# EXERCISES FOR CHILDREN WITH DYSLALIA
# -SOFTWARE INFRASTRUCTURE-

**Cristian-Eduard BELCIUG[1], Ovidiu-Andrei SCHIPOR[2], Mirela DANUBIANU[3]**
*Universitatea „Ştefan cel Mare"Suceava*
*Str. Universităţii nr.9, 720225, Suceava, Romania*
belciugc@yahoo.com, schipor@eed.usv.ro, mdanub@eed.usv.ro

***Abstract*** *In order to help children with dyslalia we created a set of software exercises. This set has a unitary software block (data base, programming language, programming philosophy). In this paper we present this software infrastructure with its advantage and disadvantage.*
*The exercises are part of a software system named LOGOMON. Therefore, besides horizontal compatibilities (between exercises) vertical compatibilities are also presented (with LOGOMON system). Concerning database tables used for modulus of exercises, a part of them is "inherited" from LOGOMON application and another are specific for exercises application. We also need to specify that there were necessary minor changes of database tables used by LOGOMON.*
*As programming language we used C#, implemented in Visual Studio 2005. We developed specific interfaces elements and classes. We also used multimedia resources that were necessary for exercises (images, correct pronouncing obtained from speech therapist recording, video clips). Another section of this application is related to loading of exercises on mobile devices (Pocket PC).*
*A part of code has been imported directly, but there were a lot of files that need to be rewritten. Anyway, the multimedia resources were used without any processing.*
***Keywords*** *exercises for dyslalia, software architecture, multimedia database, speech therapy*

# EXERCIŢII PENTRU COPIII CU DISLALIE
# -INFRASTRUCTURA SOFTWARE-

## Introducere

În articolul de faţa se vor prezenta metodele de realizare pentru modulul ce permite gestionarea cuvintelor, paronimelor, exerciţiilor şi temelor predefinite, precum şi pentru modulul ce creează teme acasă pentru un anumit copil ce se află în terapie, din temele predefinite care sunt salvate în baza de date. Tot acest modul va permite şi preluarea rezultatelor realizate de copil, rezolvând temele acasă şi salvarea lor în baza de date pentru a ţine o evidenţă a activităţii lui de-a lungul terapiei. Aceste module software au fost gândite să funcţioneze cu o bază de date, care este o parte din baza de date folosită de LOGOMON.

Interfeţele şi legătura cu baza de date au fost realizate în limbajul C# integrat in mediul de dezvoltare Visual Studio 2005.

**1. Modulul de gestiune a cuvintelor, paronimelor, exerciţiilor şi temelor predefinite**

Fiecare cuvânt care se salvează în baza de date are asociat un fişier sunet şi un fişier imagine corespunzător. Tabelele folosite în baza de date pentru acest lucru sunt: tblCuvant, tblFisierSunet, tblFisierlmagine. Tabelul tblCuvant are următoarele câmpuri:

- idCuvant - cheia primară;
- textCuvant - folosit pentru memorarea lexicografică a cuvântului respectiv;
- numeP - memorează numele persoanei care a rostit cuvântul (s-a folosit acest câmp deoarece în baza de date vor fi ţinute atât cuvintele corecte, rostite de logopezi, cât şi cuvintele rostite de pacienţi;
- prenumeP - memorează prenumele persoanei care a rostit cuvântul;
- flagLogoped - câmp folosit pentru a şti dacă persoana care a rostit cuvântul este un logoped sau un pacient;
- parteVorbire - ce parte de vorbire este cuvântul respectiv (câmpurile care urmează vor fi folosite pentru exerciţiile de recunoaştere sunet în perechi de paronime, care vor fi prezentate ulterior);
- gen - ţine genul cuvântului respectiv dacă acesta este ca parte de vorbire substantiv;

- articol - folosit pentru a şti dacă articolele "un/o" pot fi folosite/sau nu în faţa cuvântului (în exerciţiul de identificare a unui cuvânt din perechi de paronime folosind imagini; De ex: cuvântul copil poate fi folosit cu articolul "un", dar cuvântul copii nu).

Tabelele tblFisierSunet şi tblFisierlmagine au pe lângă câmpul folosit ca identificator şi câmpul cale ce memorează numele fişierului respectiv.
În baza de date este folosit tabel tblParonime, pentru memorarea perechilor de cuvinte care formează paronime. Tabelul tblParonime conţine pe lângă câmpul folosit ca identificator, şi două câmpuri pentru a memora identificatorii celor două cuvinte ce formează perechea de paronime (idCuvantl, idCuvantl).
Pentru salvarea exerciţiilor şi a configuraţiilor acestora în baza de date, se folosesc tabelele tblExercitiu şi tblConfigurare. Tabelul tblConfigurare este un tabel de legătură între tblExercitiu şi tblCuvant, deoarece acestea două se află în relaţia N:M (un cuvânt poate fi folosit în mai multe exerciţii, iar un exerciţiu poate avea mai multe cuvinte, figura 1).

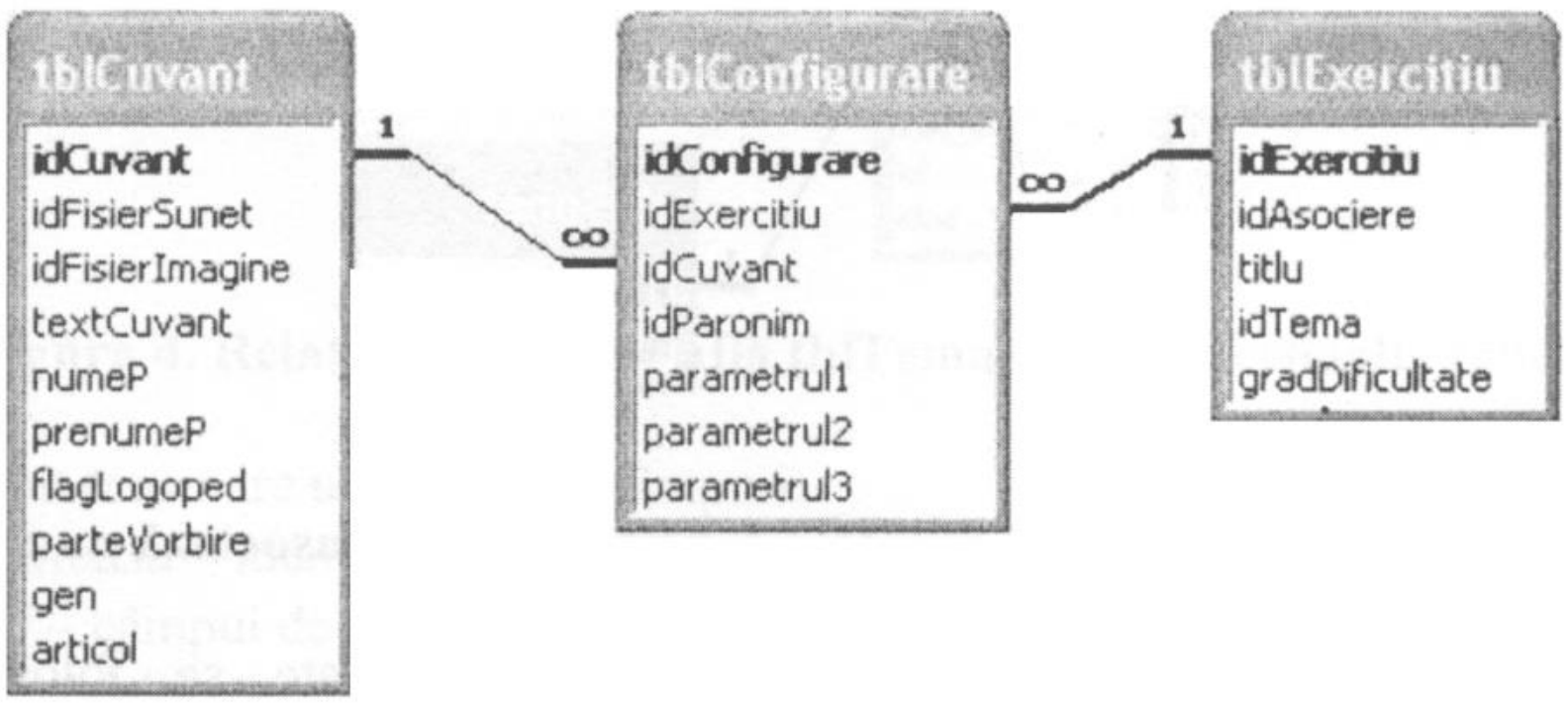

**Figura 1. Relaţia în care se află tblCuvant, tblConfigurare şi tblExercitiu**

Tabelul tblExercitiu prezintă următoarele câmpuri;

- idExercitiu - cheia primară;
- titlu - pentru a memora numele exerciţiului respectiv;
- gradDificultate - memorează gradul de dificultate al exerciţiului (un număr cuprins între 1 şi 5);
- idAsociere - câmp de legătură cu tabelele ce conţin tipul, subtipul şi sunetul vizat pentru care care este creat exerciţiul;
- idTema - câmpul de legătură pentru tabelul tblTema, ce conţine date despre indicaţiile care se dau pentru exerciţiul respectiv;

Tabelul tblConfigurare are câmpurile:

- idConfigurare — cheia primară;
- idExercitiu, idCuvant şi idParonim - sunt câmpuri de legătură cu tabelele tblExercitiu, tblCuvant şi respectiv tblParonime. Câmpul idParonim este folosit atunci când se memorează un exerciţiu de identificare cuvânt în perechi de paronime.
- pararnetrul1, parametrul2, parametru/3 - folosite pentru a memora o serie de parametri despre exerciţii, ca de exemplu: timpul cât imaginea asociată unui cuvânt este vizibilă pe ecran, dacă cuvântul are sunetul vizat sau nu, etc...

După cum se observă în figura 2, tabelul tblAsociere este folosit pentru legătura între tblExercitiu şi tblTipExercitiu, tblSubtipExercitiu şi tblSunet (relaţie N:M). Tabelele tblTipExercitiu şi tblSubtipExercitiu au pe lângă câmpul identificator şi câmpurile numeTip, respectiv numeSubtip, folosite pentru memorarea denumirii tipului/subtipului de exerciţiu, dar şi câmpurile numeAplicatie folosite pentru a memora numele aplicaţiei exerciţiului respectiv. Exemplu: Auz Fonematic este tip de exerciţiu, Identificare cuvânt în perechi de paronime este subtip de exerciţiu, Paronime.exe este numele aplicaţiei implementate pentru acest subtip de exerciţiu.

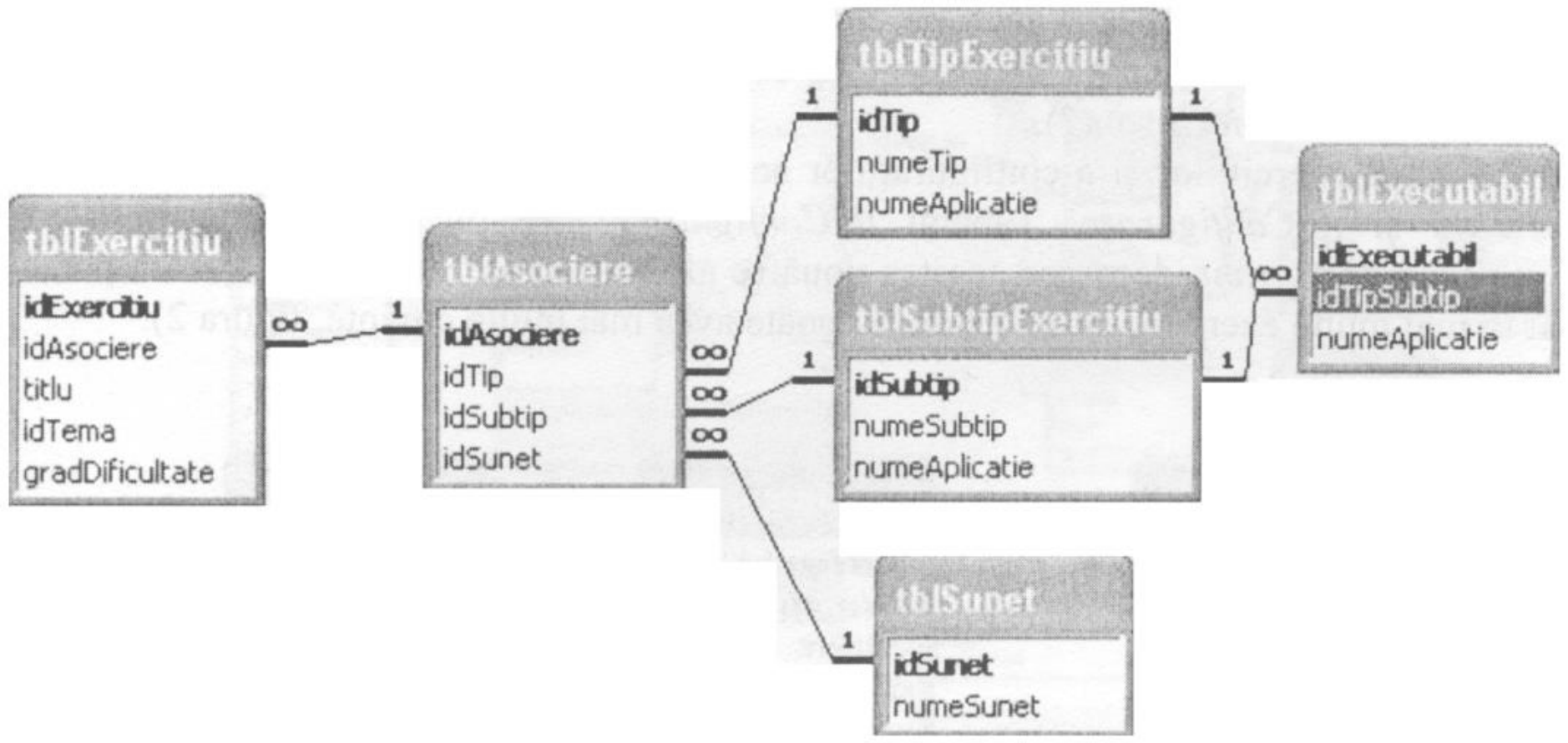

**Figura 2. Tipurile, subtipurile şi sunetele deficitare care se asociază unui exerciţiu**

Pentru memorarea unuei teme predefinite în baza de date se folosesc tabelele: tblTemaPredefinita, tblTemaPredefinitaExercitiu, Deficiente, tblTemaPredefinitaDeficienta, tblTemaPredefnitaSunet, Teste (tabelele Deficiente şi Teste sunt din baza de date a LOGOMON - ului). Tabelul tblTemaPredefinita are următoarele câmpuri:

- idTemaPredefnita - cheia primară;
- descriereTema - pentru memorarea descrierii dată pentru tema respectivă;

- numarRepetitii — memorează numărul de repetiţii pe zi recomandat pentru tema respectivă;

Tabelul tblTemaPredefinitaExercitiu este de legătură cu tabelul tblExercitiu (relaţia dintre tblTemaPredefinita şi tblExercitiu este N:M) şi prezintă pe lângă câmpul identificator (care este şi cheie primară), şi câmpul procentaj Reuşita folosit pentru memorarea procentajului ce trebuie îndeplinit pentru ca exerciţiul să fie considerat rezolvat. Tabelele tblTemaPredefinitaDeficienta şi tblTemaPredefinitaSunet sunt folosite pentru legăturile cu tabelele Deficiente şi Teste (relaţia fiind N:M).

## 2. Modulul de transmitere a temei acasă şi preluarea raportului de activitate

Tabelele din baza de date care sunt folosite de modulul de transmitere a temei acasă şi preluarea raportului de activitate sunt: tbITemaAcasa, tblTemaAcasaExercitiu. Relaţiile dintre tabele sunt prezentate in figura 3.

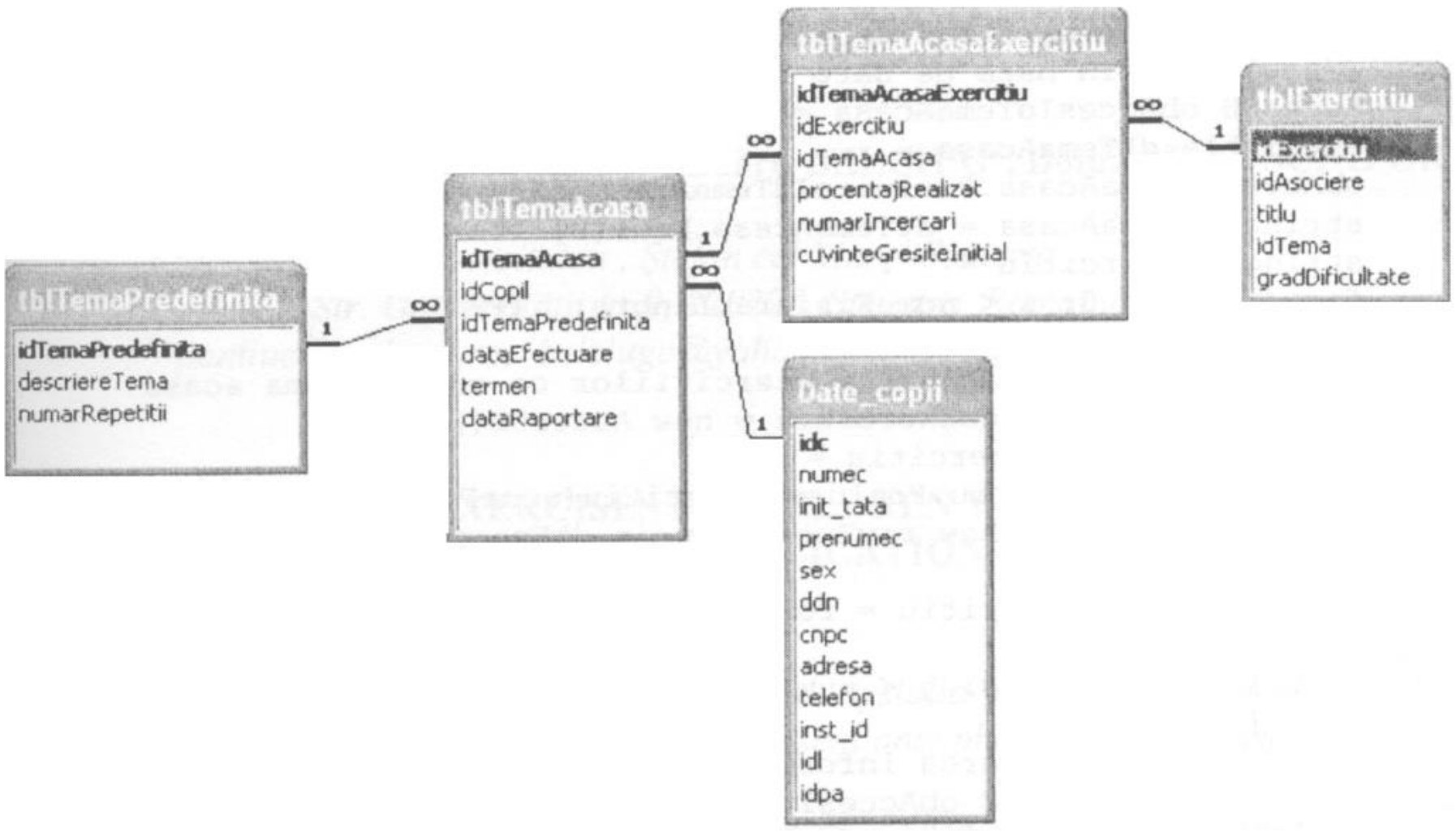

**Figura 3. Relaţiile în care se află tbITemaAcasa cu celelalte tabele**

Tabelul tblTemaAcasa are următoarele câmpuri:

- idTemaAcasa — identificatorul (cheia primară);
- idCopil - câmpul de legătură cu tabelul Date_copii (acesta păstrează datele despre copii înscrişi în terapie);
- idTemaPredefinita - câmpul de legătură cu tblTemaPredefinită;
- dataEfectuare - memorează data când a fost dată tema acasă;
- termen - termenul în zile cât are la dispoziţie logopatul să resolve tema;
- dataRaportare - acest câmp se completează atunci când copilul vine cu tema rezolvată;

Tabelul tblTemaAcasaExercitiu va fi completat integral atunci când copilul vine cu tema rezolvată. Acesta are, pe lângă câmpurile identificator şi de legătură, următoarele:

- procentaj Realizat - câmpul memorează procentajul obţinut de către copil la rezolvarea unui exerciţiu, la o anumită repetare a unui exerciţiu;
- numarlncercari - memorează numărul de repetări pentru un exerciţiu în cadrul temei respective;
- cuvinteGresitelnitial - numărul de cuvinte care au fost greşite iniţial (cuvintele greşite iniţial în cadrul unui execiţiu se vor repeta încă o dată);

## Concluzii

Dezvoltarea unui model al terapiei asistate a tulburărilor de pronunţie este un proces care trebuie tratat cu deosebită atenţie deoarece reprezintă prima etapă în cadrul dezvoltării unui sistem automat specific acestei activităţi.
În prezentul articol au fost surprinşi paşii necesari acestui demers din punct de vedere a structurii bazei de date. Au fost prezentate cele mai importante tabelele implicate împreună cu relaţiile dintre ele.

## Referinţe bibliografice

- Oster Anne-Marie, House David, Protopapas Athanassios – *Presentation of a new EU project for speech therapy: OLP (Ortho-Logo-Paedia)*, Fonetik, 2002.
- OLP, OLP Home – *URL:* http://www.xanthi.ilsp.gr/olp/default.htm, ultima modificare noiembrie 2004.
- Hatzis A., Kalikow DN, Stevens KN - *OPTICAL LOGO-THERAPY (OLT) - A computer based speech training system for the visualization of articulation using connectionist techniques*, 2002.
- Balter Olle, Engwall Olov, Oster Anne-Marie, Kjellstrom Hedvig - *ARTUR - a Computer-Based Speech Training System with Articulation Correction*, 2005.
- Eriksson Elina, Balter Olle, Engwall Olov, Oster Anne-Marie, Kjellstrom Hedvig - *Design Recommendations for a Computer-Based Speech Training System Based on End-User Interviews*, 2005.
- Bunnel H. Timothz, Yarrington M. Debra, Polikoff B. James – *STAR: Articulation Training for Young Children*, 2001.
- Viksi K., Roach P., Oster A., Kacic Z. - *A Multimedia, Multilingual Teaching and Training System for Children with Speech Disorders*, International Journal of Speech Technology, Olanda, 2000.
- Van Joolingen, W., Schreiber, A. Th., and Hermans, L. - *Handbook SDF-II*. Utrecht, The Neeherlands, CIBIT, 1999.
- Mohd Syazwan Abdullah, Andy Evans, Ian Benest, Chris Kimble - *Developing a UML Profile for Modelling KBS*, Heslington, United Kingdom, 2004.